# Generating high density molecular jets of complex neutral organic molecules using micro-sized Tesla valves

Moniruzzaman Shaikh[1,*], Xinyao Liu[1,*], Kasra Amini[1,*], Tobias Steinle[1], Jens Biegert[1,2,†]

[1]*ICFO - Institut de Ciencies Fotoniques, The Barcelona Institute of Science and Technology, 08860 Castelldefels (Barcelona), Spain.*
[2]*ICREA, Pg. Lluís Companys 23, 08010 Barcelona, Spain.*

[†]*To whom correspondence should be addressed to. Email: jens.biegert@icfo.eu.*
[*]*Authors contributed equally.*

## Abstract

We report the design and implementation of multiple tesla type micro valves in the target delivery system of a reaction microscope (ReMi) to study gas phase structural dynamics of complex polyatomic molecules, when no delivery system currently exists that can deliver dense enough molecular jets of neutral complex molecules without ionizing or exciting the target. We show, the Tesla valves provide an efficient unidirectional flow of the *cis*-stilbene molecules into the ReMi. We demonstrate using a bubbler with Tesla valves an order-of-magnitude increase in the detected stilbene molecular ion signal following the strong-field tunnel ionization (SFTI), compared to conventional bubbler without any Tesla valves. Our results for the first time, opens the door for studying large, complex neutral molecules in the gas-phase with low vapour pressures in future ultrafast studies.

## I. INTRODUCTION

Ultra-cold beams of atom and molecules[1,2] have enabled a wide range of ground-breaking experiments in physics, chemistry and biology, from capturing molecular movies of biologically relevant chemical reactions and bimolecular collisions to testing fundamental aspects of quantum mechanics.[1–14] The free expansion of gas molecules from a room temperature bottle into a high-vacuum chamber through a small pinhole generates molecular beam. This leads to a supersonic expansion of a gas to generate an internally (*i.e.* rotationally and vibrationally) cold beam of gas-phase molecules with an internal temperature of <1 K[15,16], which is a prerequisite to perform quantum state-selective studies of molecules. A further requirement is the delivery of dense enough gaseous beams to the interaction region, with typical number densities of $10^{10} - 10^{12}$ molecules/cm$^3$ required.[1,2,10,15] As the research field progressed towards the study of larger, more complex neutral gas-phase molecules, a fundamental physical challenge persists: these large molecules typically exist in the liquid or solid form at room temperature and pressure due to their relative low vapour pressure and high melting point, respectively. Delivering gas-phase jets of large neutral molecules therefore requires a phase transition from a liquid or solid to a gas. This could be achieved by: (i) heating the sample to melt a solid sample and to increase its vapour pressure; (ii) decrease the surrounding environment's pressure (*e.g.* exposing the sample to an ultra-high vacuum chamber) to decrease the vapour pressure difference between the target and its environment; or (iii) using a carrier gas (*e.g.* helium) to pass over the sample reservoir, which picks up and carries some of the vaporised molecules to the interaction region.[1,2,17] A pressure vessel containing a liquid sample in a small reservoir, called a "bubbler", that is exposed to an ultra-high vacuum can mix the vapour from the evaporating liquid sample with a carrier gas to deliver a gaseous sample to the interaction region of sufficient target density to perform measurements, as shown in FIG. 1.[17] The pressure vessel can also be heated to increase the liquid sample's vapour pressure, and in the case of a solid sample, heating the bubbler can melt the sample to form a liquid that can subsequently be evaporated to a gas. The bubbler works well for samples with a vapour pressure as low as 23 mbar at room temperature such as water.[18] Using larger target molecules with a relatively small vapour pressure (*e.g.* <1 mbar for *cis*-stilbene[19]) generates an insufficiently high gas density in the interaction region even with heating the sample to >100°C and using a helium carrier gas. It should be noted that operating at >100°C can lead to chemical degradation which should be avoided. A key and subtle limitation of this simple bubbler design is that the flow of gaseous vapour is not unidirectional which leads to the vapour becoming trapped around the sample reservoir.

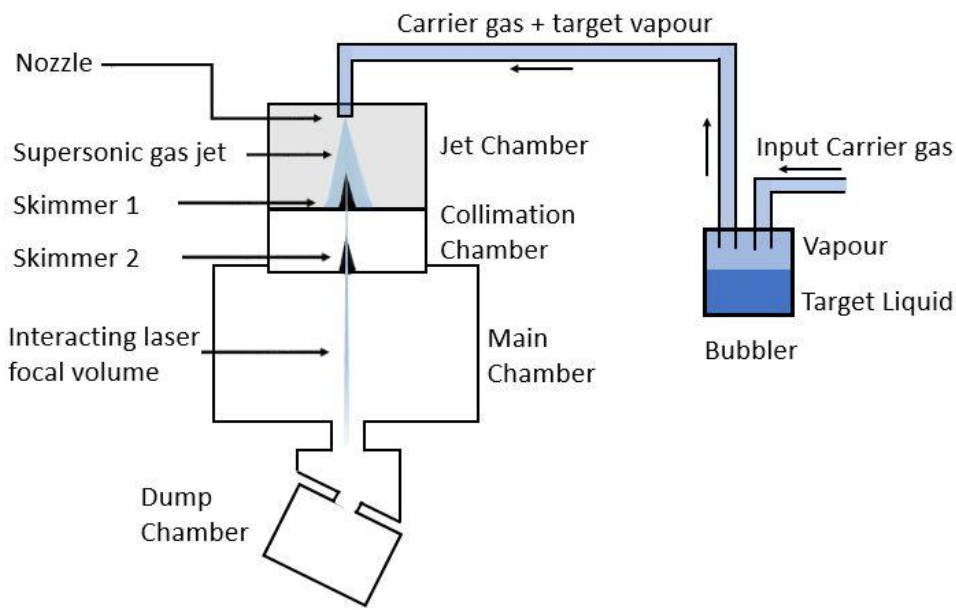

**FIG 1:** Schematic of a reaction microscope (ReMi) with typical target delivery system. A bubbler containing liquid or solid target molecule is connected and high pressure carrier gas (helium) line is connected to carry the target molecule into the ReMi.

In this paper, we present a novel bubbler design using Tesla valves[20–22], as shown in FIG. 2, that delivers a gas jet of *cis*-stilbene molecules with a unidirectional high speed flow, leading to a significantly larger gas density in the interaction-region. We demonstrate using a bubbler with Tesla valves an order-of-magnitude increase in the detected stilbene molecular ion signal following the strong-field tunnel ionization (SFTI)[9,23,24] of stilbene molecules in the presence of a 97 fs (FWHM) 3.2 µm laser pulse. We show that SFTI studies can be subsequently performed at lower peak laser intensities with Tesla valves, minimizing the fragmentation of large molecules which typically possess low ionization potentials ($I_p$) and are more susceptible to fragmentation. This is particularly important for SFTI techniques such as laser-induced electron diffraction (LIED) of intact molecular ions where the fragmentation of the molecule must be avoided.[9,23]

This paper has been divided into five major sections: In Section **II**, the working principle of a Tesla valve is discussed. Section **III** describes the experimental setup. The experimental results are shown in section in Sec. **IV**. Discussions and conclusion are described in Sec **V**.

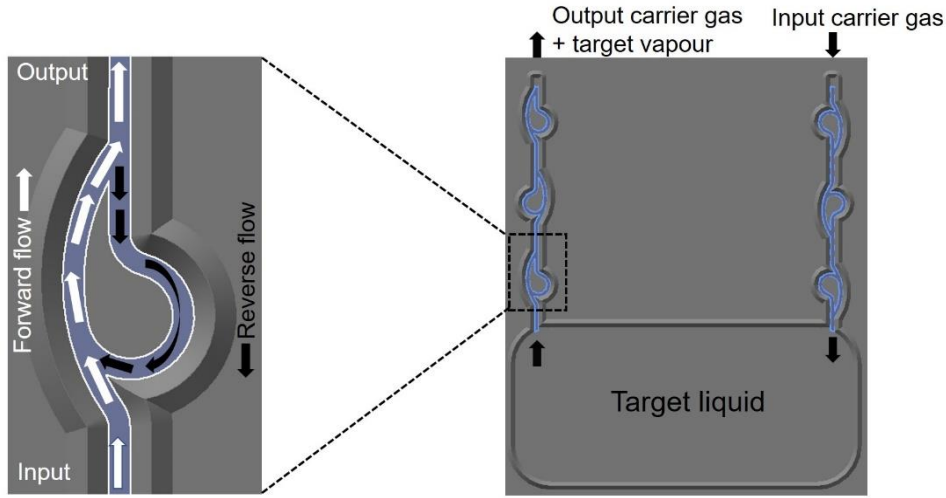

**FIG. 2.** (Left) Working principle of a Tesla valve. White and black arrows indicate the forward and the reverse flow respectively. (Right) Bubbler with the Tesla valves, where the channel of the valves are in shaded blue in colour. The valves in the input line and the output line are arranged such that there is effective unidirectional flow of carrier gas and efficient transport of the target vapour through the bubbler output.

## II. WORKING PRINCIPLE OF A TESLA VALVE

Tesla valves are no-moving-parts check valves which ensure a unidirectional flow by allowing a fluid or gas to preferentially flow in one direction without obstruction, while providing an impermeable barrier to its flow in the reverse direction, as shown in FIG. 2. Tesla valves typically consist of two in-parallel channels comprising of a curved bifurcating channel, called a helix region, and a straight-channel portion, called the trunk region, as shown in FIG. 2. Tesla valves are easy to manufacture in micro sized and require no external power for operation.[25] The effectiveness of Tesla valves in achieving unidirectional flow is characterized by the pressure diodicity parameter, $D_{i,p}$, given by

$$D_{i,p} = \frac{\Delta p_r}{\Delta p_f},$$

where achieving a unidirectional flow requires a higher pressure drop in the reverse flow direction, $\Delta p_r$, than in forward flow direction, $\Delta p_f$, with typical $D_{i,p}$ values of 1.5 and above achievable.[25–27] A further advantage of achieving a unidirectional flow is the ability to avoid the flow of the target sample in the reverse direction to ensure the stable delivery of the maximum sample vapour to the interaction region.

## III. EXPERIMENTAL SETUP

Our experimental set-up has been described in detail elsewhere (see Refs[6,23]), with a brief description given here. We use two types of bubbler reservoir setups, as shown in FIG. 1 and FIG. 2.: (i) a standard rectangular bubbler without Tesla valves consisting of a reservoir (50 mL) and two inlet/outlet metallic tubes for the transport of helium carrier gas and the gas-phase target molecule; and (ii) a rectangular bubbler with Tesla valves containing a small reservoir (10 mL) and two inlet/out tubes, respectively. In both cases, the vaporized gas-phase target molecules are supersonically expanded into a vacuum chamber and subsequently collimated using two skimmer stages. The collimated molecular beam interacts with an orthogonally directed 97 fs (FWHM) 3.2 µm linearly polarized laser pulse with a pulse energy of 6.25 − 11.25 µJ generated from a home-built optical parametric chirped pulse amplifier (OPCPA) setup with up to 21 W output power at a repetition rate of 160 kHz.[28,29] An on-axis paraboloid placed inside of the reaction microscope (ReMi)[30] focuses the 3.2 µm laser pulse into the molecular beam down to a focal spot of 6 − 7 µm. We performed measurements at a variety of peak pulse intensities ranging between $8.7 \times 10^{13}$ – $1.6 \times 10^{14}$ Wcm$^{-2}$ corresponding to a Keldysh parameter[31] of $\gamma \approx 0.3$. Homogeneous electric and magnetic extraction fields (12.5 Vcm$^{-1}$ and 10.5 G, respectively) guide the resulting ions and electrons towards separate detectors, each of which consist of chevron-stacked dual microchannel plates coupled with a quad delay-line anode set-up. The resulting particles are detected in full coincidence with 4π acceptance.

## IV. EXPERIMENTAL RESULTS

We present ion data measured with a ReMi[30] following the STFI of gas-phase *cis*-stilbene using a bubbler with and without Tesla valves. FIG. 3 shows the *y*-position hit on the ion detector for each detected ion as a function of its time-of-flight (ToF) measured using a bubbler with and without Tesla valves. The corresponding ion time-of-flight (ToF) spectra measured using a bubbler with (black line) and without (red line) Tesla valves following the SFTI of *cis*-stilbene molecules at $1.5 - 1.6 \times 10^{14}$ Wcm$^{-2}$ are shown in FIG. 3. In the presence of a strong laser field, the stilbene molecule with a relatively low ionization potential, $I_p$, of 7.80 eV[33] can either: (i) lose an electron to form a *cis*-stilbene molecular ion ($C_{14}H_{12}^+$; ToF peak at 15.6 µs); or (ii) lose two or more electrons to Coulomb explode and generate two or more ion fragments (*e.g.* $C_7H_6^+ + C_7H_6^+$; ToF peaks at around 11.0 µs). A comparison of the measured $C_{14}H_{12}^+$ molecular ion signal using a bubbler with and without Tesla valves is shown through the zoom-in of the $C_{14}H_{12}^+$ ToF signal in FIG. 4. The contribution of the molecular ion signal to the full spectra is significantly increased by a factor of 6.5 with the use of Tesla valves. Moreover, we find that the stilbene molecular ion constitutes 2.94% (0.45%) of all detected ion signal with (without) using Tesla valves. Simultaneously, the contribution of ion fragments such as $C_7H_6^+$ and $C_{10}H_8^+$ are also increased with the use of Tesla valves, confirming that a denser molecular jet is delivered to the interaction region. It is also important to note the length of time that the *cis*-stilbene molecular beam was stable for. We observed that a bubble without Tesla valves provided ion signal for approximately five minutes (*i.e.* at a repetition rate of 160 kHz this corresponds to 48 million laser shots), which is approximately two orders of magnitude too-short-a-timescale typically required to perform, for example, LIED measurements. We attribute the appearance of signal from the saturated vapour of stilbene in the reservoir which already formed whilst placing the sample in the reservoir, with no further stilbene vapour produced when exposed to the ultra-high vacuum chamber. Our Tesla valve bubbler in fact bypassed this problem as we observed that using only 1.5 of *cis*-stilbene liquid sample, we were able to measure *cis*-stilbene ion data for as long as six hours (*i.e.* 3,456 million laser shots) without any problems. Thus, we assign the observed problem experienced in the bubbler without Tesla valves to the high speed non-unidirectional flow of stilbene vapour, which we attribute to the sample vapour being stuck and localized around the bubbler reservoir after 10 minutes and thus leading to the inability to transport any of the target to the interaction region inside the ReMi. Thus, the use of Tesla valves provides favourable conditions to (i) deliver a molecular jet of *cis*-stilbene molecules to the interaction region over a relatively long measurement time, and (ii) to achieve a sufficiently large enough molecular ion signal that forms a large proportion ($>5\%$) of the total detected ion signal.

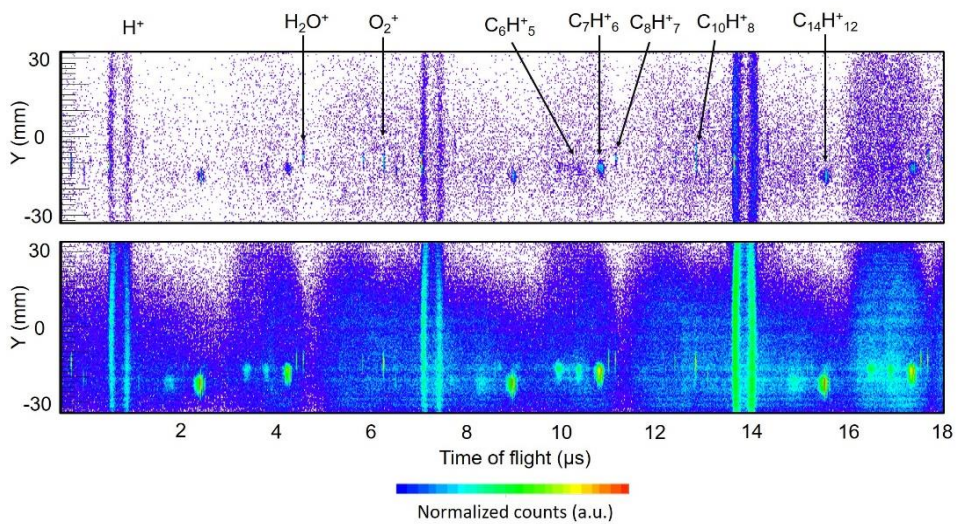

**FIG. 3.** (Top) Y vs time of flight (ToF) for the case of a bubbler without Tesla valves and (Bttom) bubbler with Tesla valves respectively.

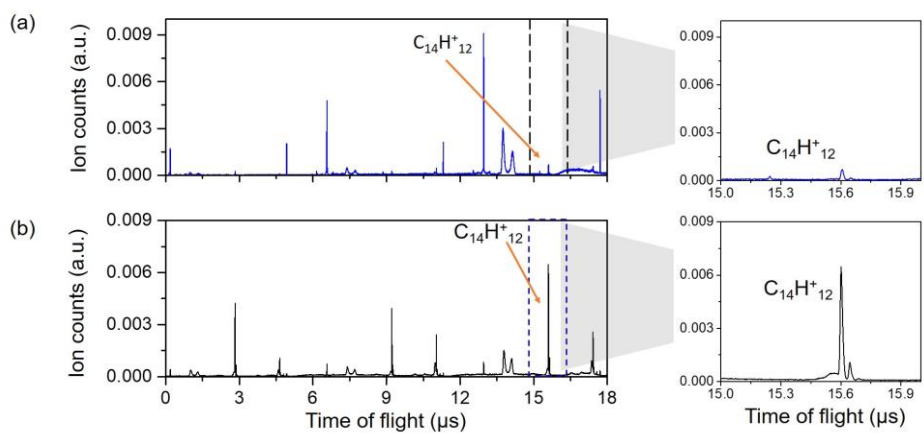

**FIG. 4**. Ion time of flight (ToF) for the case of (a) bubbler without Tesla valves and (b) bubbler with Tesla valves. Respective stilbene ion ($C_{14}H^+_{12}$) peaks (15.6 µs) in the ToFs are identified and shown on the right side.

After establishing the advantages of using Tesla valves in a bubbler, we next focus on a systematic study of the bubbler's dependence on laser power and helium carrier gas pressure. FIG. 5 shows the ion ToF spectra measured with a Tesla valve bubbler at a laser power of 1.8 W (black line) and 1.0 W (red line). Generally in strong-field physics, if the molecular ion is of interest (such as in LIED measurements of the molecular ion) then the peak laser intensity, $I_0$ (which is a function of laser power, pulse duration and focal spot size) must be kept at a minimum in order to avoid fragmenting the molecular ion but at the same time $I_0$ must be high enough to generate appreciable ion signal for detection. We demonstrate that reducing the laser power from the highest laser power safe enough for our detector to perform measurements (1.8 W) to the lowest laser power where appreciable ion signal is still detectable and sufficient for measurement (1.0 W), a 32% increase in the contribution of the $C_{14}H_{12}^+$ molecular ion signal to the total detected ion signal is observed (from 2.9% to 4.0%, respectively – see FIG. 5). This leads to an improved ~9% of the molecular ion signal contribution to the total signal when compared to the corresponding value for a bubbler without Tesla valves. At the same time, a similar magnitude decrease in the contribution of ion fragments to the total ion signal is simultaneously observed, confirming that a more stable, undissociated molecular ion is generated at 1.0 W. We next investigate the dependence of the ion signal from a Tesla valve bubbler on the helium carrier gas backing pressure between 490 – 900 mbar of helium, with the corresponding ToF spectra shown in FIG. 6. We find that the molecular ion signal increases by ~20% when increasing the helium pressure from 490 mbar to 900 mbar, as shown in FIG. 6. Just as in the laser power dependence earlier, a similar order-of-magnitude decrease in the contribution of ion fragments to the total ion signal is simultaneously seen.

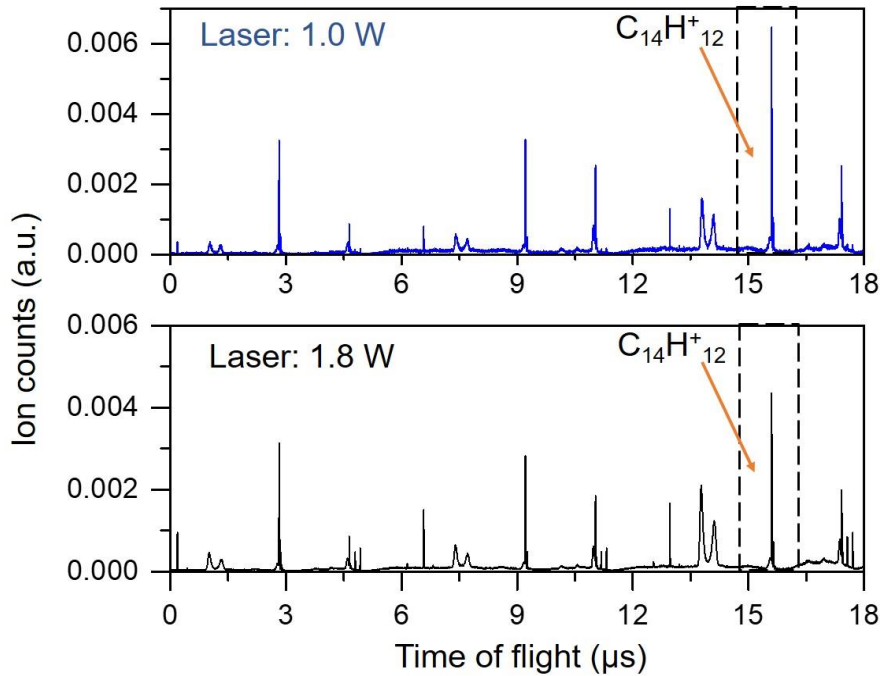

**FIG. 5.** Ion time of flight (ToF) as a function of laser intensity. Stilbene ion peaks (15.6 µs) in the ToFs are identified within the dotted rectangular boxes.

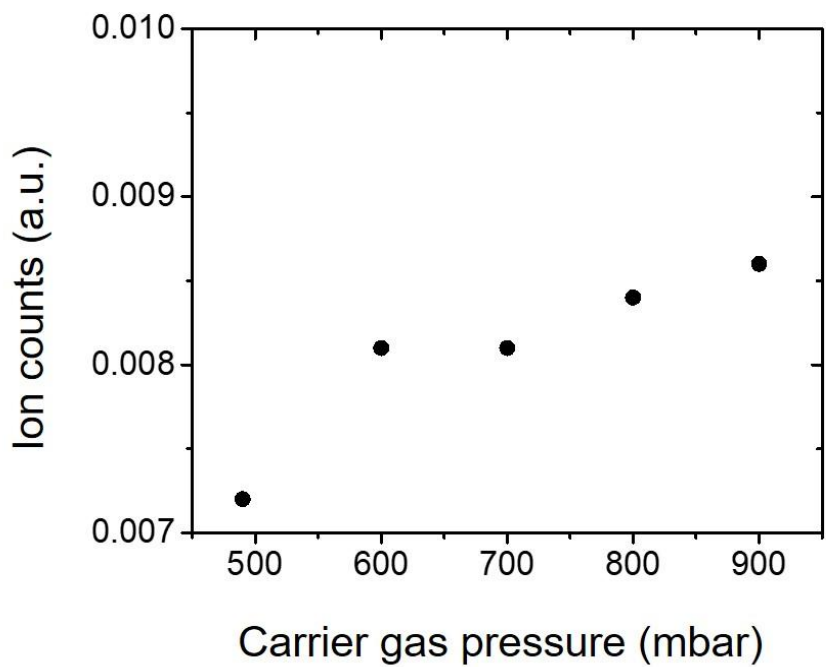

**FIG. 6.** Stilbene ion counts in the total ion signal as a function of carrier gas backing pressure.

## V. DISCUSSIONS AND CONCLUSION

In this paper, we have demonstrated the capability to deliver a molecular jet of gas-phase *cis*-stilbene molecules over a relatively long period of measurement time (> 6 hours), making it possible, for the first time, which will enable future STFI measurements of such large and complex molecules. We achieve this by using a bubbler delivery system with Tesla valves that ensures the unidirectional flow of *cis*-stilbene vapour to the interaction region which would have otherwise been stuck and localized around the bubbler reservoir in the absence of Tesla valves. Secondly, we achieve an order-of-magnitude increase in the molecular ion contribution to the total detected ion signal with the use of Tesla valves, increasing the quality and signal-to-noise of our *cis*-stilbene gas-phase measurements to ultimately minimize measurement time. We systematically characterized the dependence of our ion signal from our Tesla valve bubbler on experimental conditions such as the peak laser intensity and helium carrier gas backing pressure. We show that the molecular ion signal can be maximized by performing measurements at: (i) a low peak laser intensity; and (ii) high helium gas backing pressure. Our flow simulations quantitatively reveal the velocity and pressure distributions of gas flow through the Tesla valve bubbler, confirming the fast and unidirectional flow of gas flow that is achieved using Tesla valves. We believe that our new bubbler design utilizing Tesla valves opens the pathway towards the study of large, complex gas-phase neutral molecules that are internally cold enough to perform cold, ultracold and ultrafast measurements. We believe that the Tesla valve bubbler will become an important gas delivery system in time-resolved ultrafast electron diffraction pump-probe measurements of large, complex molecules (*e.g.* stilbene, azobenzene).

**Acknowledgements**
J.B. and group acknowledge financial support from the European Research Council for ERC Advanced Grant "TRANSFORMER" (788218), ERC Proof of Concept Grant "miniX" (840010), FET-OPEN "PETACom" (829153), FET-OPEN "OPTOlogic" (899794), Laserlab-Europe (EU-H2020 654148), MINECO for Plan Nacional FIS2017-89536-P; AGAUR for 2017 SGR 1639, MINECO for "Severo Ochoa" (SEV- 2015-0522), Fundació Cellex Barcelona, CERCA Programme / Generalitat de Catalunya, the Polish National Science Center within the project Symfonia, 2016/20/W/ST4/00314, and the Alexander von Humboldt Foundation for the Friedrich Wilhelm Bessel Prize. J.B. and K.A. acknowledge the Polish National Science Center within the project Symfonia, 2016/20/W/ST4/00314. X.L. and J.B. acknowledge financial support from China Scholarship Council. We acknowledge help and support from Xavi Menino.